\documentclass[prb,superscriptaddress,floatfix,twocolumn,showpacs,amsfonts,amsmath,amssymb]{revtex4}
\usepackage{graphicx} 
\usepackage{dcolumn} 
\usepackage{bm} 

\begin{document}

\title{Effective action of the weakly doped $t$-$J$ model and spin-wave excitations in the spin-glass phase of La$_{2-x}$Sr$_x$CuO$_4$}

\author{Andreas L\"uscher}
\affiliation{Institut Romand de Recherche Num\'erique en Physique des Mat\'eriaux (IRRMA), EPFL, 1015 Lausanne, Switzerland}
\author{Alexander I. Milstein}
\affiliation{Budker Institute of Nuclear Physics, 630090 Novosibirsk, Russia}
\author{Oleg P. Sushkov}
\affiliation{School of Physics, University of New South Wales, Sydney 2052, Australia}

\date{\today}

\begin{abstract}
We derive the low-energy effective field theory of the extended $t$-$J$ model in the regime of light doping. The action includes a previously unknown Berry phase term, which is discussed in detail. We use this effective field theory to calculate spin-wave excitations in the disordered spin spiral state of La$_{2-x}$Sr$_x$CuO$_4$ (the spin-glass phase). We predict an excitation spectrum with two distinct branches: The in-plane mode has the usual linear spin-wave dispersion and the out-of-plane mode shows non-trivial doping dependent features. We also calculate the intensities for inelastic neutron scattering in these modes.
\end{abstract}


\maketitle

\section{Introduction}
The phase diagram of the prototypical cuprate superconductor La$_{2-x}$Sr$_x$CuO$_4$  (LSCO) shows that the magnetic state changes tremendously with Sr doping. The three-dimensional antiferromagnetic (AF) N\'eel order identified~\cite{keimer92} below $325\ \text{K}$ in the parent compound disappears at doping $x\approx0.02$ and gives way to the so-called spin-glass phase which extends up to $x\approx 0.055$. In both, the N\'eel and the spin-glass phase, the system essentially behaves as an Anderson insulator and exhibits only hopping conductivity. Superconductivity then sets in for doping $x\gtrsim 0.055$, see Ref.~\onlinecite{kastner98}.  One of the most intriguing properties of LSCO is the static incommensurate magnetic ordering observed at low temperature in \emph{elastic} neutron scattering  experiments. This ordering manifests itself as a scattering peak shifted with respect to the  antiferromagnetic position. Very importantly, the incommensurate ordering is a generic feature of LSCO. According to experiments in the N\'eel phase, the incommensurability is almost doping independent and  directed along the orthorhombic $b$ axis~\cite{matsuda02}. In the spin-glass phase, the shift is also directed along the $b$ axis, but scales linearly with doping~\cite{wakimoto99,matsuda00,fujita02}. Finally, in the underdoped superconducting region ($0.055 \lesssim x \lesssim 0.12$), the shift still scales linearly with doping, but it is directed along the crystal axes of the tetragonal lattice~\cite{yamada98}. 
Apart from measurements at $x=0.024$ reported in Ref.~\onlinecite{matsuda00},
 all \emph{inelastic} studies on LSCO have so far been performed at sufficiently high 
doping~\cite{thurston89,mason92,cheong91}, where the material is conducting and the 
incommensurate structures are thus always aligned along the Cu-O bonds. Similar incommensurate features have also been observed in inelastic neutron scattering in YBCO, see, e.g., Refs.~\onlinecite{bourges00,fong00,mook02,hayden04,hinkov04,stock06}, or Ref.~\onlinecite{tranquada05} for a review. 

The observation of these static incommensurate peaks in LSCO caused a renewal of theoretical interest in the idea of spin spirals in cuprates~\cite{hasselmann04,sushkov04,juricic04,sushkov05,lindgard05,juricic06}. While the static spiral in the conducting state ($x \gtrsim 0.055$) is probably an oversimplification, we strongly believe that it represents the right picture of insulating LSCO ($x \lesssim 0.055$), where mobile holes are trapped in the vicinity of Sr dopants. Based on a fully controlled solution of the $t$-$J$ model, we first analyzed the low-doping regime ($x \lesssim 0.02$) and demonstrated~\cite{luscher06} that local static spirals are present in the N\'eel state. In a recent paper~\cite{luscher07}, we extended this model to the spin-glass phase of LSCO ($0.02 \lesssim x \lesssim 0.055$) and found that the ground state is not really a spin glass, but a disordered spin spiral state. To prevent confusion, we nevertheless comply with the usual terminology and refer to this regime as the spin-glass phase. The structure factors obtained in our approach are in perfect agreement with elastic neutron scattering experiments.

The effective Lagrangian for the weakly doped $t$-$J$ model has been derived quite some time ago~\cite{wiegman88,wen89,shraiman88}. The kinematic structure of the static limit established in Ref.~\onlinecite{shraiman88} can be used without modifications to calculate the above mentioned ground state properties of LSCO~\cite{luscher06,luscher07}, the only difference being the material dependent parameters. However, the derivation of the time-dependent terms, necessary to describe excitations, is non-trivial and we thus discuss it in detail in the present work. The rest of the paper is organized as follows:  In Sec.~\ref{sec:model}, we derive the general effective field theory describing the low-energy sector of the the lightly doped $t$-$J$ model. This effective action applies to the uniformly doped conducting state as well as the disordered insulating state. In Sec.~\ref{sec:results}, the effective model is used to calculate the spectrum of spin-wave excitations and the neutron scattering cross sections in insulating LSCO. These predictions can be directly verified in inelastic neutron scattering experiments. Our conclusions are presented in Sec.~\ref{sec:conclusion}.

\section{Effective low-energy action of lightly doped two-dimensional $t-t'-t''-J$ model\label{sec:model}}
Over a decade ago, the two-dimensional $t$-$J$ model has been suggested to describe the essential low-energy physics of high-$T_c$ cuprates~\cite{anderson87,emery87,zhang88}. In its extended version, this model includes additional hopping matrix elements $t'$ and $t''$ to next-nearest neighbors. The Hamiltonian of the model is well known, see, e.g., Ref.~\onlinecite{sushkov04}, and we do not present it here. The numerical values of the parameters of the $t$-$t'$-$t''$-$J$ model for LSCO follow from Raman spectroscopy~\cite{tokura90} and ab-initio calculations~\cite{andersen95}. It is convenient to measure energies in units of $J \approx 130 \ \text{meV}$, so that $t=3.1$, $t'=-0.6$ and $t''=0.3$. At zero doping (no holes), the $t$-$J$ model is equivalent to the Heisenberg model and describes the Mott insulator La$_2$CuO$_4$ (LCO). The removal of a single electron from this Mott insulator, or in other words the injection of a hole, allows the charge carrier to propagate. Single-hole properties of the $t$-$J$ model are well understood~\cite{dagotto94}. The main features are a very flat dispersion along the edges of the magnetic Brillouin zone (MBZ) with four degenerate half-pockets centered at $S=\left(\pm\frac{\pi}{2},\pm\frac{\pi}{2}\right)$.  The quasi-particle residue $Z$  at the minimum of the dispersion is $Z\approx0.3$. In the full-pocket description, where two half-pockets are shifted  by the AF vector ${\bf Q}_{AF}=\left(\pi,\pi\right)$, the two minima are located at $S_a=\left(\frac{\pi}{2},\frac{\pi}{2}\right)$ and $S_b=\left(\frac{\pi}{2},-\frac{\pi}{2}\right)$. The system is thus somewhat similar to a two-valley semiconductor.

\begin{figure*}
\includegraphics[width=0.8\textwidth,clip]{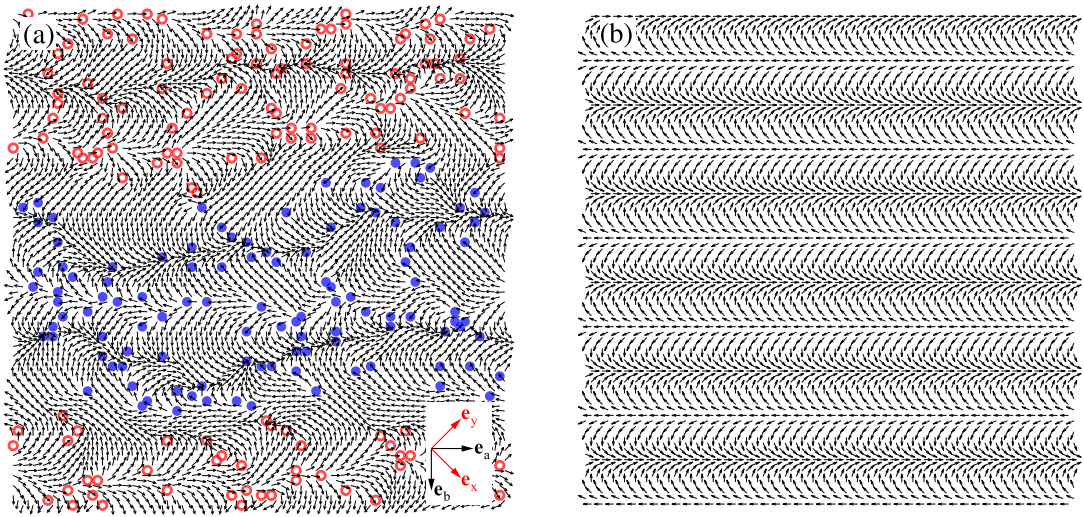}
\caption{(Color online). (a) Characteristic ground state configuration of a particular realization at doping $x=0.05$. The impurity pseudospins $\frac{1}{2}\Psi_i{\vec \sigma}\Psi_{i}$ (circles) are oriented along the $z$ axis. Full (open) circles correspond to values $-\frac{1}{2}$ ($+\frac{1}{2}$). Small arrows represent the ${\vec n}$-field. The system forms domains stretched along the $a$ direction, in which all the pseudospins are aligned in parallel. (b) Uniform spiral state~(\ref{gs0}) with average pitch~(\ref{q}). Apart from the absence of disorder, a uniform spiral is very similar to the true ground state and thus a good starting point to describe excitations. The corresponding structure factors are shown in Fig.~\ref{fig:structurefactor}.} 
\label{fig:groundstate}
\end{figure*}

\subsection{Static limit}
The relevant energy scale for small uniform doping at zero temperature is of the order of $xJ \ll J$. In the case of nonuniform doping, when holes are trapped near Sr ions in hydrogenlike bound states, the relevant energy scale is the binding energy of the hole, which is about $10\ \text{meV}$, see Ref.~\onlinecite{sushkov05}. Since in both situations, the energy scales are much smaller than $J$, one can simplify the Hamiltonian of the $t$-$J$ model by integrating out all high-energy fluctuations. For the hole-doped case, the effective energy density reads
\begin{multline} \label{E0}
{\cal E}=
\frac{\rho_s}{2}\left({\bf \nabla}{\vec n}\right)^2
+\sum_{\alpha} \left\{Ê\psi^{\dag}_{\alpha}
\left[\frac{\beta{\bf \nabla}^2}{2}+\Delta_{\alpha}+V({\bf r})\right]\psi_{\alpha} 
\right. \\ \left. 
- \sqrt{2}g (\psi^{\dag}_{\alpha}{\vec \sigma}\psi_{\alpha})
\cdot\left[{\vec n} \times ({\bf e}_{\alpha}\cdot{\bf \nabla}){\vec n}\right]\right\} \ .
\end{multline}
Here ${\vec n}({\bf r})$ is the staggered component of the copper spins and 
$\rho_s \approx 0.18J$ is the spin stiffness.  $\psi_{\alpha}({\bf r})$ is a fermionic spinor operator describing the holes, with  an index $\alpha=a,b$ (flavor) indicating the  location of the hole in momentum space (either in pocket $S_{a}$ or $S_{b}$). The operator ${\vec \sigma}$ is a psudospin that originates from the existence of two  sublattices at zero doping and ${\bf e}_{\alpha}=(1/\sqrt{2},\pm 1/\sqrt{2})$ is a unit  vector orthogonal to the face of the MBZ where the hole is located. The first term in 
Eq.~(\ref{E0}) is the usual elastic energy of the nonlinear $\sigma$ model (NLSM) and the 
last term is the interaction of the hole with the twist of the ${\vec n}$-field that 
favors the formation of local spirals. The kinematic structure of this term was first 
derived in Ref.~\onlinecite{shraiman88}, and the coupling constant $g \approx Zt\approx J$ 
was calculated in Ref.~\onlinecite{igarashi92}. The second term represents the 
single-particle energy of a hole. The kinetic energy is expanded around the center of the 
corresponding hole pocket. In general, the kinetic energy is anisotropic. However, for 
the values of $t'$, $t''$, and $t$ relevant for LSCO, the dispersion is practically 
isotropic with $\beta\approx 2J$, see Ref.~\onlinecite{sushkov04}. Note that we set the 
lattice spacing of the square lattice equal to unity. Substituting the physical values for
 the lattice spacing, one can check that  $\beta\approx 2J$ corresponds to an effective 
mass of about two electron masses. The bottom of the hole band for a given flavor is equal
 to $\Delta_{\alpha}$ and we have the freedom to set $\Delta_b = 0$. For the tetragonal 
structure, the hole pockets $S_{a}$ and $S_{b}$ are degenerate and therefore 
$\Delta_a=\Delta_b=0$. 
This degeneracy is lifted by the orthorhombic distortion in the low-temperatire phase
 of LSCO. The difference in energy is simply due to the slightly different distances 
between neighboring sites along the orthorhombic $a$ and $b$ directions. According to a 
recent calculation~\cite{jepsen}, $\Delta_a \sim 7\ \text{meV}$. In the spin-glass phase 
of LSCO, holes therefore only occupy the $S_{b}$ hole pocket, which in turn explains the 
pinning of the incommensurate diagonal spin structure to the orthorhombic $b$ axis, see 
Ref.~\onlinecite{luscher06}. Finally, $V({\bf r})$ is the Coulomb potential due to the Sr
 ions that leads to hydrogenlike bound states. In the effective energy~(\ref{E0}), we 
have neglected the Dzyaloshinskii-Moriya and the $XY$ anisotropies, see 
Ref.~\onlinecite{chovan00asilvaneto05a}, because these anisotropies are small and 
therefore not important for purpose of the present work. A more formal discussion of the 
effective energy~(\ref{E0}) can be found in Ref.~\onlinecite{wiese05}. 
We note that the pure $t$-$J$ model ($t'=t''=0$) is unstable with respect to
phase separation and/or short-range charge stripe formation~\cite{sushkov04}. In this 
case, the long-wavelength approximation considered in this section is meaningless. However, the extended $t$-$t'$-$t''$-$J$ model at physical values of $t'$ and $t''$ is stable in the charge sector~\cite{sushkov04} and the long-wavelength action is thus well defined.

\subsection{Derivation of the time-dependent Lagrangian\label{sec:berryphase}}
The static limit, i.e., the effective energy given by Eq.~(\ref{E0}), is sufficient to 
determine ground state properties of insulating LSCO~\cite{luscher06,luscher07}.
In order to calculate excitation spectra, one has to include the time dependence of the 
staggered field ${\vec n}$ and the hole operator $\psi$. Na\"{\i}vely adding the usual 
time derivatives, we find the Lagrangian
\begin{equation} \label{eq:Lprime}
{\cal L}'=\frac{\chi_{\perp}}{2}{\dot{\vec n}}^2+
\sum_{\alpha}\left\{ \frac{i}{2}
(\psi^{\dag}_{\alpha}{\dot \psi}_{\alpha}-
{\dot \psi}^{\dag}_{\alpha}\psi_{\alpha})
\right\}-{\cal E} \ ,
\end{equation}
where $\chi_{\perp}\approx 0.5/(8J)$ is the perpendicular magnetic susceptibility and ${\cal E}$ is given by Eq.~(\ref{E0}). However, due to the interaction of the holes with the ${\vec n}$-field [last term in Eq.~(\ref{E0})], there is an additional non-trivial term. Let us find its kinematic structure by following the original calculation of the static limit~\cite{shraiman88} and extend it to the time-dependent case: The hopping term between nearest sites $i$ and $j$ in the $t$-$J$ model is given by
\begin{equation*} 
H_t=-t\sum_{\left\langle ij\right\rangle,s} c_{is}^{\dag}c_{js} + \text{H.c.}\ ,
\end{equation*}
where $c_{is}^{\dag}$ creates an electron on site $i$ with spin projection $s=\pm1/2$ onto a globally fixed $z$ axis. For the N\'eel state with a single injected hole $\left|h_{\xi\alpha}\right\rangle$ on the $\xi=\uparrow,\downarrow$ sublattice, with flavor index $\alpha=a,b$, the expectation value $\left\langle h_{\uparrow\alpha} \right| H_{t} \left| h_{\downarrow\alpha} \right\rangle$ vanishes because spins on nearest sites are exactly opposite.
 Let us now assume that the spin background is deformed in such a way that the staggered 
magnetization is given by
 ${\vec n}=(\sin\theta\cos\varphi,\sin\theta\sin\varphi,\cos\theta)$ where 
$\theta$ and $\varphi$ smoothly depend on ${\bf r}$. Because we assume the deformation 
of the background to be smooth, the two sublattices $\uparrow$ and $\downarrow$ are still
 well defined, but the arrows no longer represent the projection of the spin along the  $z$-axis.
After the unitary transformation of the operators $c_{is}$ into the local 
reference frame with quantization axis $z$ directed along ${\vec n}$, we find 
\begin{equation} \label{Htt}
H_t \to-\sqrt{2}t\left(h_{\uparrow\alpha}^{\dag}({\bf r})[\theta'+i\varphi'\sin\theta]
h_{\downarrow\alpha}({\bf r})+\text{H.c.}\right) \ ,
\end{equation}
where $F'=({\bf e}_\alpha\cdot{\bf \nabla})F$. Due to high-energy quantum 
fluctuations, $t$ is replaced by $t \to g=Zt$, see Ref.~\onlinecite{igarashi92}. The Hamiltonian~(\ref{Htt}) leads to the following Schr\"odinger equation for the two-component wave function $v_{\alpha}=\left(h_{\uparrow\alpha}, h_{\downarrow\alpha}\right)^T$
\begin{equation} \label{Sv}
i \ \dot v_{\alpha}=-\sqrt{2}g({\vec f}\cdot{\vec \sigma})v_{\alpha} \ ,
\end{equation}
where ${\vec f}=(\theta',\varphi'\sin\theta, 0)$ and ${\vec \sigma}$ are the Pauli matrices. Since 
\begin{equation*}
f^2=(\theta')^2+(\varphi')^2\sin^2\theta= \left[{\vec n} \times ({\bf e}_{\alpha}\cdot{\bf \nabla}){\vec n}\right]^2 \ ,
\end{equation*}
one can perform a unitary transformation $U$, $\psi_{\alpha}=U v_{\alpha}$, such that 
\begin{equation*}
U^{\dag}({\vec f}\cdot{\vec \sigma})U=\left[{\vec n} \times ({\bf e}_{\alpha}\cdot{\bf \nabla}){\vec n}\right]\cdot{\vec \sigma} \ .
\end{equation*}
The Schr\"odinger equation~(\ref{Sv}) for a single hole is thus transformed into
\begin{equation} \label{Spsi}
i\ \dot{\psi}_{\alpha}=-\sqrt{2}g
\left[{\vec n} \times ({\bf e}_{\alpha}\cdot{\bf \nabla}){\vec n}\right]\cdot
{\vec \sigma}\psi_{\alpha}
-U^{\dag}\dot{U}\psi_{\alpha} \ .
\end{equation}
If the ${\vec n}$-field is static, then $\dot{U}=0$ and the Schr\"odinger equation~(\ref{Spsi}) is equivalent to the Euler-Lagrange equation that follows from the effective action~(\ref{eq:Lprime}). Taking into account the $\dot{U}$ contribution in~(\ref{Spsi}) thus leads to the introduction of an additional term in the effective action. Its kinematic structure can be determined due to the following reasons: 1) The term originates from $U^{\dag}{\dot U}$ and thus contains only one time derivative ${\dot {\vec n}}$. 2) Because it acts on a pseudospinor, it contains ${\vec \sigma}$. 3) It must be a scalar. 4) It is independent of the coupling constant $g$. 
The only structure that satisfies these conditions is 
$\psi_{\alpha}^\dag{\vec \sigma}\psi_{\alpha} \cdot[{\vec n}\times{\dot {\vec n}}]$. 
A calculation shows that it appears with a universal coefficient $-1/2$ and  therefore 
leads to the Lagrangian
\begin{multline} \label{L00}
{\cal L}=
\frac{\chi_{\perp}}{2}{\dot{\vec n}}^2+
 \sum_{\alpha} \left\{ \frac{i}{2}
(\psi^{\dag}_{\alpha}{\dot \psi}_{\alpha}-
{\dot \psi}^{\dag}_{\alpha}\psi_{\alpha}) \right. \\
\left. -\frac{1}{2}\left(\psi_{\alpha}^\dag {\vec \sigma} \psi_{\alpha}\right) 
\cdot[{\vec n}\times{\dot {\vec n}}] \right\}-{\cal E} \ .
\end{multline}
The additional ${\vec \sigma}\cdot[{\vec n}\times{\dot {\vec n}}]$ term, closely related to the Berry phase, has been obtained before for Kondo-lattice-like models~\cite{kubert93,sikkema97}. However, in these models the term appears with a non-universal coefficient that depends on the parameters of the microscopic description. It is important to note that the effective Lagrangian~(\ref{L00}) is not restricted to the situation of localized holes, but is also valid in the uniformly doped case when $V({\bf r})$ in Eq.~(\ref{E0}) is zero.

\begin{figure*}
\includegraphics[width=0.6\textwidth,clip]{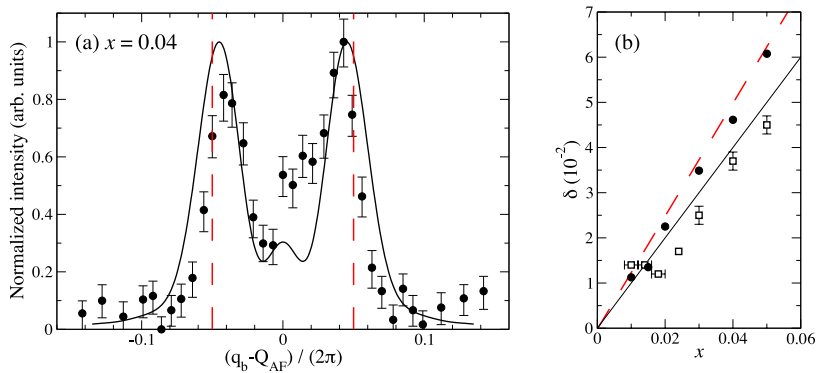}
\caption{(Color online). (a) Neutron scattering probability $S_{\bf q}$ for doping $x=0.04$. Full circles correspond to experimental observations taken from Fig.~4 in Ref.~\onlinecite{fujita02}, with normalized intensities. The solid line represents our simulation~\cite{luscher07}, containing no fitting parameters. The dashed line indicates the positions of the delta function peaks obtained from a uniform spiral. (b) Incommensurability $\delta$ (half the distance between the two peaks measured in reciprocal lattice units of the tetragonal lattice), as a function of doping. Accurate Monte Carlo calculations (dots) are in agreement with experimental data (squares) taken from Refs.~\onlinecite{wakimoto00a,matsuda00,matsuda02}, see Fig. 6 of Ref.~\onlinecite{matsuda02}. The uniform spiral (dashed line) captures the essential properties of the spin-glass ground state.} \label{fig:structurefactor}
\end{figure*}

\subsection{Larmor's theorem}
The validity of the prefactor $-\frac{1}{2}$ in the Berry phase term can be verified in a situation where 
the behavior of the system is well known: According to Larmor's theorem, spins in a uniform magnetic field 
${\vec B}$ precess with frequency ${\vec \omega}={\vec B}$. In the undoped case, this property follows 
directly from the Euler-Lagrange equations. According to Refs.~\onlinecite{chakravarty88,fisher89}, the 
Lagrangian of the NLSM in an external magnetic field is given by
\begin{equation*} 
{\cal L}_B=\frac{\chi_{\perp}}{2}\left({\dot {\vec n}}-[{\vec n}\times{\vec B}]\right)^2-
\frac{\rho_s}{2}[{\bf \nabla}{\vec n}]^2 \ .
\end{equation*}
Note that we set $g_s\mu_B \rightarrow 1$, $g_s$ being the gyromagnetic factor and $\mu_{B}$ the Bohr 
magneton. 
For a uniform field ${\bf \nabla}{\vec n}=0$, the Euler-Lagrange equation 
${\dot{\vec n}}=[{\vec n}\times{\vec B}]$ describes the precession of the staggered field around ${\vec B}$, 
in accordance with Larmor's theorem. 

In a doped system, the situation is similar. Let us for simplicity consider a single hole trapped by the Sr potential $V({\bf r})$ and omit the flavor index $\alpha$. The generalization to multiple dopants is 
straightforward. Using the pseudospin density
\begin{equation*}
{\vec \xi}({\bf r})=\psi^{\dag}({\bf r}){\vec \sigma}\psi ({\bf r}) \ ,
\end{equation*}
we perform the variation of Eq.~(\ref{L00}) with respect to ${\vec n}({\bf r})$ and find
\begin{equation} \label{static}
\Delta{\vec n}+\frac{\sqrt{2}g}{\rho_s}\left([{\vec n}\times{\vec \xi'}]+ 2[{\vec n'}\times{\vec  \xi}]\right)
+\lambda{\vec n}=0 \ .
\end{equation}
Here $F'=({\bf e}_{\alpha}\cdot{\bf \nabla})F$ and $\lambda=\lambda({\bf r})$ is a Lagrange multiplier taking into account the constraint ${\vec n}^2=1$. It has been shown in Ref.~\onlinecite{sushkov05} that 
${\vec n}({\bf r})$ that results from Eq.~(\ref{static}) lies in a plane to which the pseudospin is 
perpendicular. In other words, ${\vec \xi}$ is parallel to ${\vec n} \times {\vec n'}$.

If we place the doped system in a uniform magnetic field, we have to replace
\begin{equation*}
{\dot {\vec n}}^2 \to \left({\dot {\vec n}}-[{\vec n}\times{\vec B}]\right)^2
\end{equation*}
in the Lagrangian~(\ref{L00}), and we also have to take into account the interaction energy of the pseudospin with the magnetic field, ${\cal E}_B=-\frac{1}{2}({\vec \xi}\cdot{\vec n})({\vec B}\cdot{\vec n})$, that has been derived in Ref.~\onlinecite{luscher06}. In the presence of a uniform magnetic field, the effective Lagrangian thus reads
\begin{multline} \label{LB1}
{\cal L}_B=
\frac{\chi_{\perp}}{2}\left({\dot {\vec n}}-[{\vec n}\times{\vec B}]\right)^2
-\frac{\rho_s}{2}\left({\bf \nabla}{\vec n}\right)^2 \\
+\frac{i}{2}(\psi^{\dag}{\dot \psi}-{\dot \psi}^{\dag}\psi) 
-\psi^{\dag}\left[\frac{\beta{\bf \nabla}^2}{2}+V(r)\right]\psi \\
-\frac{1}{2}{\vec \xi}\cdot[{\vec  n}\times {\dot {\vec n}}]
+\sqrt{2}g {\vec \xi}\cdot\left[{\vec n} \times {\vec n'}\right]
+\frac{1}{2}({\vec \xi}\cdot{\vec n})({\vec B}\cdot{\vec n}) \ .
\end{multline}
Here we have omitted the long derivative ${\bf p} \to {\bf p} -\frac{e}{c}{\bf A}$, describing the interaction of the magnetic field with the electric charge, because we are only interested in the spin dynamics. Larmor's theorem implies that ${\dot{\vec n}}={\vec n}\times{\vec B}$ still satisfies the equation of motion. Substituting this solution into Eq.~(\ref{LB1}), we find
\begin{multline} \label{Lxi}
{\cal L}_B \to 
-\frac{\rho_s}{2}\left[{\bf \nabla}{\vec n}\right]^2
+\frac{i}{2}(\psi^{\dag}{\dot \psi}-{\dot \psi}^{\dag}\psi)
-\psi^{\dag}\left[\frac{\beta {\bf \nabla}^2}{2}+V(r)\right]\psi \\
+ \sqrt{2}g {\vec \xi}\cdot\left[{\vec n} \times {\vec n'}\right]
+\frac{1}{2}({\vec \xi}\cdot{\vec B})
 \ .
\end{multline}
Since according to Eq.~(\ref{static}), ${\vec \xi}$ is parallel to ${\vec n} \times {\vec n'}$, the equation of motion that follows from the Lagrangian~(\ref{Lxi}) is
\begin{equation} \label{xit}
{\dot {\vec \xi}}={\vec \xi}\times{\vec B} \ ,
\end{equation}
i.e., ${\vec \xi}$ also precesses with frequency ${\vec \omega}={\vec B}$. Eq.~(\ref{static}) therefore remains valid in the proper reference frame and we conclude that the Berry phase term is crucially important for the fulfillment of Larmor's theorem.

\section{Spin-wave excitations in the spin-glass phase of LSCO\label{sec:results}}
In the N\'eel and the spin-glass phase, holes are localized near Sr ions in hydrogenlike bound states~\cite{sushkov05}. In what follows, we refer to these bound states as impurities. The binding energy is about 10 meV and the description is therefore valid at temperatures well below 100 K. A direct experimental indication of this picture is the variable-range hopping conductivity observed in insulating LSCO, see Ref.~\onlinecite{ando02}, or Ref.~\onlinecite{kastner98} for a review. 

\subsection{Uniform spiral approximation for the ground state}
In the case of hole localization, the hole wave function $\psi_{i}({\bf r})$ decays exponentially away from the Sr ion located at position ${\bf R}_{i}$
\begin{equation*}
\psi_{i}({\bf r}) = \Psi_{i} \sqrt{\frac{2}{\pi}} \kappa e^{-\kappa \left|{\bf r}-{\bf R}_{i}\right|} \ .
\end{equation*}
The energy~(\ref{E0}) is then given by
\begin{equation}
\label{e00}
{\cal E}=
\frac{\rho_s}{2}\left[{\bf \nabla}{\vec n}\right]^2
- \sqrt{2}g \sum_i \rho\left({\bf r-R_i}\right)
 \Psi^\dag_i {\vec \sigma} \Psi_i
\left[{\vec n} \times \left({\bf e}_{b}\cdot{\bf \nabla}\right){\vec n}\right] \ ,
\end{equation}
with $\rho\left({\bf r}\right)=\frac{2}{\pi}\kappa^2e^{-2\kappa r}$ and the
 inverse localization length $\kappa \sim 0.4$ that follows form hopping conductivity~\cite{kastner98}. 
We remind the reader that we set the lattice spacing of the tetragonal structure equal to unity.  In absolute units, the hole localization length is thus approximately 10 \AA.
 Note that the pseudospinor $\Psi_i$ of a given impurity $i$ is independent of
 ${\bf r}$. We do not include the binding energy in (\ref{e00})
 because it just shifts the energy by a constant. The ground state is found by 
minimizing the energy~(\ref{e00}). After integrating out the ${\vec n}$-field, 
this minimization procedure reduces to finding the minimum of a system of 
randomly distributed interacting dipoles 
${\vec \xi}_{i} = \Psi_{i} {\vec \sigma} \Psi_{i}$. This task has been 
accomplished in our previous work~\cite{luscher07} using Monte Carlo 
simulations. A characteristic ground state configuration of the dipoles and 
the ${\vec n}$-field obtained for a particular realization of Sr positions 
at average doping $x=0.05$ is shown in Fig.~\ref{fig:groundstate}(a). 
We would  like to emphasize that this state is not really a spin-glass, but a 
disordered  spin spiral state~\cite{luscher07}. The spiral structure manifests
itself as  incommensurate peaks in the structure factor 
\begin{equation*}
S_{\bf q} \propto \sum_{ij,\alpha} e^{i {\bf q} \cdot \left( {\bf r}_{i}-{\bf r}_{j}\right)} n^\alpha\left({\bf r}_{i}\right) n^\alpha\left({\bf r}_{j}\right) \ .
\end{equation*}
Our claim about the disordered spiral state is supported by experimental 
observations: While the incommensurate structure is already observed below 
$30-40 \ \text{K}$, the irreversible glassy behavior only sets in below 
$T\sim 5-6 \ \text{K}$~\cite{niedermayer98awakimoto00b} and is in our opinion 
due to the inter-layer interaction that leads to a freezing of incompatible 
spiral configurations in different planes.
The profile of the neutron scattering cross section calculated in 
Ref.~\onlinecite{luscher07} is shown in Fig.~\ref{fig:structurefactor}(a), 
together with experimental data at doping $x=0.04$. Our results are in good 
agreement with experiments, especially given the fact that they contain no 
fitting parameters. Even though the values for the important parameters 
$\kappa$ and $g$ are fixed, there is some variability related to their 
accuracy. In the present work, as well as our two recent 
papers~\cite{luscher06,luscher07}, we use $\kappa=0.4$ and $g\approx1$. 
However, we believe that $\kappa=0.3$ or $g=0.8$ are well possible. The 
spiral has an average pitch proportional to doping, as can be seen in the 
doping dependence of the incommensurability $\delta$, defined as half the 
distance between the two peaks, shown in Fig.~\ref{fig:structurefactor}(b).

The observed broadening of the incommensurate peaks is on the one hand due to intrinsic disorder caused by the random distribution of the Sr ions and on the other hand due to spin vortices (topological defects) that determine the size of the domains with a given chirality of the spiral, see Ref.~\onlinecite{luscher07}. These broadening mechanisms can only be taken into account in numerical studies. However, if we neglect these effects, we can find the ground state in a much simpler mean-field approach: Since the interaction term in Eq.~(\ref{e00}) is maximal when all the pseudospins are orthogonal to plane of the spiral, we can take
\begin{align} \label{gs0}
{\vec \xi}_i&=\Psi_i{\vec \sigma}\Psi_i=(0,0,1)\ ,\nonumber\\
{\vec n}&=(\cos{\bf Q\cdot r},\sin{\bf Q\cdot r},0) \ .
\end{align}
After averaging over the impurity positions, we find the energy ${\cal E}=\rho_s/2 Q^2-\sqrt{2}gx({\bf e}_b\cdot {\bf Q})$.
The spiral pitch that minimizes this energy is given by 
\begin{equation}
\label{q} {\bf Q}=\frac{\sqrt{2}g x}{\rho_s}{\bf e}_b \ ,
\end{equation}
and agrees well with accurate Monte Carlo simulations and experimental data, see Fig.~\ref{fig:groundstate}, Fig.~\ref{fig:structurefactor} and Ref.~\onlinecite{luscher07}. Hence, apart from the absence of disorder, a uniform spiral~(\ref{gs0}) with average pitch~(\ref{q}) is very similar to the true ground state and is thus a good starting point to describe excitations.

\begin{figure}
\includegraphics[width=0.45\textwidth,clip]{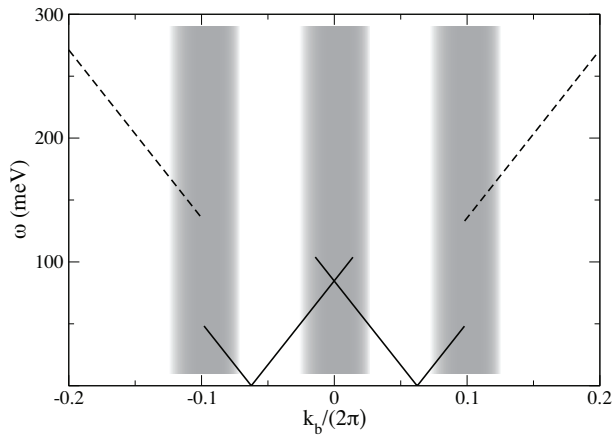}
\caption{Spectrum of in-plane spin waves excited in neutron scattering at doping $x=0.05$. The momentum is directed along  the orthorhombic $b$ direction and the incommensurable vector~(\ref{q}) is $Q\approx 0.39$. There are two linear branches that start at $k_b=\pm Q$, that are broadened due to disorder and ultimately disappear at $|{\bf k}\pm{\bf Q}|\sim \sqrt{x}$. For larger momenta, spin waves reappear with usual dispersion of the N\'eel antiferromagnet. The frequency at the intersection ${\bf k}_b=0$ scales linerly with doping.}\label{fig:inplane}
\end{figure}

\subsection{In-plane spin-wave excitations}
An in-plane excitation is described by a small deviation $\varphi=\varphi(t,{\bf r})$ from the uniform spiral ground state~(\ref{gs0}). Substituting the expression for the ${\vec n}$-field
\begin{equation}
\label{inp}
{\vec n}=(\cos({\bf Q\cdot r}+\varphi),\sin({\bf Q\cdot r}+\varphi),0) \ .
\end{equation}
into the effective Lagrangian~(\ref{L00}), we find
\begin{multline}
\label{Lphi1}
{\cal L} =\frac{\chi_{\perp}}{2}{\dot \varphi}^2-
\frac{\rho_s}{2}(Q^2+2Q\varphi'+(\nabla\varphi)^2) \\
+\sum_i\rho({\bf r}-{\bf R}_i)\left\{\sqrt{2}g\xi_i(Q+\varphi')
-\frac{1}{2}\xi_i{\dot \varphi} \right. \\ \left. 
+\frac{i}{2}(\Psi^{\dag}_i{\dot \Psi}_i-{\dot \Psi}^{\dag}_i\Psi_i)\right\}\ ,
\end{multline}
where $\xi_i=\Psi_i^\dag\sigma_z\Psi_i$. It should be clearly understood that Eq.~(\ref{inp}) and the Lagrangian~(\ref{Lphi1}) are only valid in the long-wavelength limit, $q \lesssim \sqrt{x}$,
where $q$ is the typical momentum of the $\varphi$-field and $\sqrt{x}$ is the inverse average distance between impurities. In this limit, one can average over impurity positions and replace $\sum_i\rho({\bf r}-{\bf R}_i) \to x$, which leads to the cancellation of terms proportional to $\varphi'$ in the Lagrangian~(\ref{Lphi1}) because of Eq.~(\ref{q}). From the Euler-Lagrange equation for $\Psi$, we find ${\dot \xi}=0$. The equation of motion for $\varphi$ therefore reads
\begin{align*} 
\chi_{\perp}{\ddot\varphi}=\rho_s\nabla^2\varphi \ .
\end{align*}
The spectrum of in-plane excitations is thus exactly the same as the spin-wave spectrum in undoped LCO, $\omega =cq$, with the spin-wave velocity $c=\sqrt{ \rho_{s}/\chi_{\perp}}\approx1.66$. The corresponding Green's function of the $\varphi$-field is given by
\begin{equation*} 
G_{in}=\frac{\chi_{\perp}^{-1}}{\omega^2-c^2q^2+i0} \ ,
\end{equation*}
where the subscript {\it in} stays for in-plane.

Let us also calculate the neutron scattering cross section, in order to make specific predictions for future experiments. The Hamiltonian describing the interaction of the neutron spin ${\vec S}^N$ with the ${\vec n}$-field reads
\begin{equation} \label{n}
H^N\propto {\vec S}^N\cdot{\vec n}=S^N_z n_z+\frac{1}{2}\left(S^N_+n_-+S^N_-n_+\right) \ .
\end{equation}
After the substitution of the in-plane excitation~(\ref{inp}), the above Hamiltonian reads
\begin{align*} 
H^N&\propto\frac{1}{2}S^N_+e^{-i({\bf Q}\cdot{\bf r}+\varphi)}
+\frac{1}{2}S^N_-e^{i({\bf Q}\cdot{\bf r}+\varphi)} \nonumber \\
&\to\frac{1}{2}e^{i{\bf k}\cdot{\bf r}}\left\{S^N_+e^{-i{\bf Q}\cdot{\bf r}}(1-i\varphi)
+S^N_-e^{i{\bf Q}\cdot{\bf r}}(1+i\varphi)\right\} \ ,
\end{align*}
where ${\bf k}$ is the momentum transfer and ${\bf Q}$ the momentum shift due to the spiral ground state. The scattering probability for unpolarized neutrons is given by
\begin{eqnarray} 
\label{i1}
&&I_{in}(\omega,{\bf k}) \propto
-\frac{1}{8\pi}\text{Im}\left[G_{in}(\omega,{\bf k}-{\bf Q})
+G_{in}(\omega,{\bf k}+{\bf Q})\right] \nonumber\\
&&=\frac{1}{16\chi_{\perp}\omega}\left[\delta(\omega-c|{\bf k}-{\bf Q}|)
+\delta(\omega-c|{\bf k}+{\bf Q}|)\right] \ .
\end{eqnarray}
Fig.~\ref{fig:inplane} shows the spectrum of excited spin waves for a momentum transfer ${\bf k}=(0,k_b)$ chosen along the orthorhombic $b$ direction at doping $x=0.05$. Very importantly, the above calculation is valid at $q=|{\bf k}\pm{\bf Q}| \lesssim \sqrt{x}$ and in the absence of disorder. In experiments, we expect a broadening similar to that obtained in elastic neutron scattering, see Fig.~\ref{fig:structurefactor}. For momenta $q \sim \sqrt{x}$ (shaded area in Fig.~\ref{fig:inplane}), the peaks disappear because of very strong broadening due to scattering of the spin-wave off random impurities. However, we predict that at even higher momenta, the peaks reappear as usual spin waves, $\omega=ck$, with some broadening, reflecting the nearly perfect antiferromagnetic alignment of nearest neighbor spins.

\begin{figure}
\includegraphics[width=0.35\textwidth,clip]{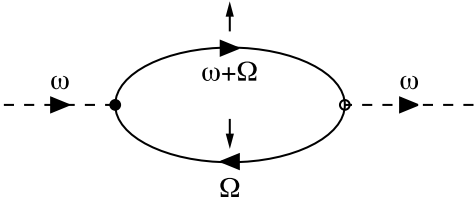}
\caption{Polarization operator describing the scattering of a spin-wave (dashed line) off a pseudospin (solid line). The corresponding pesudospin-flip vertices (circles) are given by Eq.~(\ref{g12}).}\label{fig:polarization}
\end{figure}

\subsection{Out-of-plane spin-wave excitation}
An out-of-plane excitation can be represented as a small deviation $n^z\left({\bf r},t\right)$ from the uniform spiral ground state~(\ref{gs0}). Let us write the ${\vec n}$-field as 
\begin{equation*} 
{\vec n}=(\sqrt{1-n_z^2}\cos{\bf Q}\cdot{\bf r}, \sqrt{1-n_z^2}\sin{\bf Q}\cdot{\bf r}, n_z)\ ,
\end{equation*}
and substitute this expression into the effective Lagrangian~(\ref{L00}). Neglecting cubic and higher order terms in $n_z$, we decompose the Lagrangian into two parts ${\cal L}={\cal L}_0+{\cal L}_\text{int}$, with
\begin{widetext}
\begin{align} \label{Lnzz}
{\cal L}_0&=\frac{\chi_{\perp}}{2} {\dot n_z}^2
-\frac{\rho_s}{2}\left\{-Q^2+({\bf \nabla}n_z)^2+Q^2n_z^2\right\}+
\sum_i\left[\frac{i}{2}(\Psi^{\dag}_i{\dot \Psi}_i-{\dot \Psi}^{\dag}_i\Psi_i)
-\frac{\Delta}{2}\Psi_i^{\dag}(1-\sigma_z)\Psi_i\right] \ ,\nonumber\\
{\cal L}_\text{int}&= -\sum_i\rho({\bf r}-{\bf R}_i) \Psi^{\dag}_i\left\{ \sigma_+e^{-i{\bf Q}\cdot{\bf r}}\left[\sqrt{2}g(Qn_z-in_z')+\frac{i}{2}{\dot  n}_z\right]+\sigma_-e^{i{\bf Q}\cdot{\bf r}}\left[\sqrt{2}g(Qn_z+in_z')-\frac{i}{2}{\dot n}_z\right] \right\} \Psi_i\ .
\end{align}
\end{widetext}
Here $\Delta=2\sqrt{2}gQ=4g^2x/\rho_s$ is the energy required to flip a pseudospin and $F'=({\bf e}_b\cdot{\bf \nabla})F$. From the diagonal part of the Lagrangian ${\cal L}_0$, we find the unperturbed spin-wave Green's function
\begin{equation} \label{g0}
G_0=\frac{\chi_{\perp}^{-1}}{\omega^2-\omega_{\bf q}^2} \ ,
\end{equation}
where $\omega_{\bf q}^2=c^2(q^2+Q^2)$. Note that this is \emph{different} from the usual spin-wave dispersion. In order to calculate the corrections to the bare propagator due to ${\cal L}_\text{int}$, we introduce the second quantization representation for the ${\vec n}$-field
\begin{equation*} 
n_z({\bf r})=\sum_{\bf q}\frac{1}{\sqrt{2\chi_{\perp}\omega_{\bf q}}}\left(
e^{i\omega_{\bf q}t-i{\bf q}\cdot{\bf r}}m_{\bf q}^{\dag}+
e^{-i\omega_{\bf q}t+i{\bf q}\cdot{\bf r}}m_{\bf q}\right) \ ,
\end{equation*}
with the spin-wave creation and annihilation operators $m_{\bf q}^{\dag}$ and $m_{\bf q}$. The off-diagonal part of the Lagrangian ${\cal L}_\text{int}$~(\ref{Lnzz}) gives rise to the pseudospin-flip vertices 
\begin{align} \label{g12}
\Gamma_{\uparrow\downarrow}&=\frac{1}{\sqrt{\chi_{\perp}}}
e^{i({\bf Q}+{\bf q})\cdot{\bf R}_i}F({\bf Q}+{\bf q})
\left[\sqrt{2}g(Q-q_b)-\frac{\omega}{2}\right] \ , \nonumber\\
\Gamma_{\downarrow\uparrow}&=\frac{1}{\sqrt{\chi_{\perp}}}
e^{i({\bf Q}-{\bf q})\cdot{\bf R}_i}F({\bf Q}-{\bf q})
\left[\sqrt{2}g(Q+q_b)+\frac{\omega}{2}\right] \ .
\end{align}
In contrast to the in-plane excitations, the Berry phase term derived in Sec.~\ref{sec:berryphase} is important for the out-of-plane spectrum because it leads to the additional terms $\pm \omega/2$ in the above expression. The impurity form factor $F({\bf p})$ is given by
\begin{align} \label{F}
F({\bf p})&=\int d^2r \ \rho({\bf r})e^{i{\bf p}\cdot{\bf r}}
= \frac{2\kappa^2}{\pi} \int d^2r e^{-2\kappa r}e^{i{\bf p}\cdot{\bf r}} \nonumber \\&=\frac{1}{\left[1+p^2/(4\kappa^2)\right]^{3/2}}\approx 1-\frac{3}{8}\frac{p^2}{\kappa^2} \ ,
\end{align}
where for the last expression, we have used the small momentum limit $p \ll \kappa$. 

The polarizability of a given impurity is described by the diagram shown in Fig.~\ref{fig:polarization}.
Using the bare Green's function~(\ref{g0}) together with the pesudospin-flip vertices~(\ref{g12}), the polarization operator is given by the polarizability of a single impurity multiplied by the concentration $x$
\begin{multline*} 
P(\omega,{\bf q})=
x\left\{\frac{\left|{\Gamma_{\uparrow \downarrow}}\right|^2}{\omega-\Delta}+
\frac{\left|{\Gamma_{\downarrow\uparrow}}\right|^2}{-\omega-\Delta}\right\} = \\
\frac{c^2\Delta}{2}Ê\left\{
F^2({\bf Q}+{\bf q})H(-\omega,-{\bf q})+
F^2({\bf Q}-{\bf  q})H(\omega,{\bf q})
\right\} \ ,
\end{multline*}
where 
\begin{equation*}
H(\omega,{\bf q}) = - \frac{\left(Q+q_b+Q\omega/\Delta\right)^2}{\omega+\Delta} \ .
\end{equation*}
The polarization operator is invariant under a simultaneous change of the signs of ${\bf q}$ and $\omega$, i.e., $P(\omega,{\bf q})=P(-\omega,-{\bf q})$. However, it is neither an even function of the frequency $\omega$ nor the momentum ${\bf q}$. In order to find the spectrum for small momenta, we substitute the approximate expansion for the form factor~(\ref{F}) and neglect quartic and higher order terms in ${\bf q}$. In this limit, the small asymmetry disappears and we find
\begin{multline} \label{p11}
P(\omega,{\bf q})\approx -c^2\left\{Q^2+\frac{q_b^2}{1-\omega^2/\Delta^2}\right. \\ \left.- \frac{3}{4\kappa^2} \left[(Q^2-q_b^2)^2+q_a^2(Q^2+q_b^2)\right]\right\} \ .
\end{multline}
Including single-loop corrections, the Green's function of the $n_z$-field reads
\begin{equation*} 
G_{out}=\frac{\chi_{\perp}^{-1}}{\omega^2-\omega_{\bf q}^2-P(\omega,{\bf q})} \ ,
\end{equation*}
where the subscript {\it out} stays for out-plane. The excitation spectrum is given by poles of the Green's function. Using the symmetric expression for the {polarization operator}~(\ref{p11}), we find two branches
\begin{align*}
\Omega_{1,{\bf q}}&\approx c\sqrt{
\frac{q_a^2+\frac{3}{4\kappa^2}\left[(Q^2-q_b^2)^2+q_a^2(Q^2+q_b^2)\right]}
{1+\frac{c^2q_b^2}{\Delta^2}}} \nonumber \ ,\\
\Omega_{2,{\bf q}}&\approx \Delta+\frac{c^2q_b^2}{2\Delta} \ .
\end{align*}

\begin{figure}
\includegraphics[width=0.45\textwidth,clip]{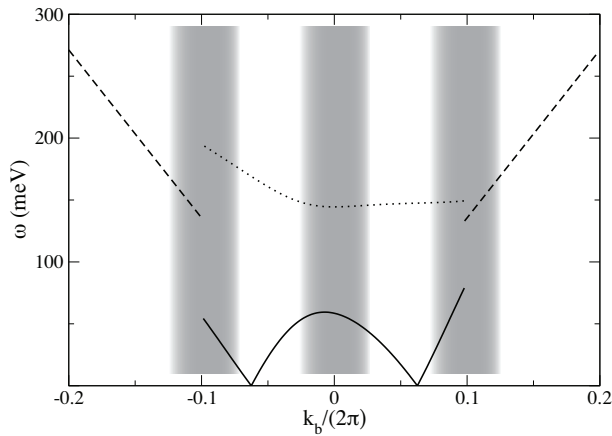}
\caption{Spectrum of out-of-plane spin-wave excitations at doping $x=0.05$. The Momentum is directed
 along the orthorhombic $b$ direction. The inverse radius of the impurity wave-function is $\kappa=0.4$
 and the incommensurable vector~(\ref{q}) is $Q\approx 0.39$. Both, the lower and upper branch are 
getting broad and ultimately disappear due to  disorder at $|{\bf k}\pm{\bf Q}|\sim \kappa$. The upper
 branch has a very small spectral weight and it thus difficult to observe in neutron scattering. 
At larger momenta, the spin waves reappear with usual dispersion of the N\'eel antiferromagnet. The frequency at the top of the dome at ${\bf k}_b=0$ scales quadratically with doping.}
\label{fig:outofplane}
\end{figure}

The lower branch vanishes at ${\bf q}=\pm {\bf Q}$, in accordance with the Goldstone theorem, while the upper branch has a gap $\Delta$. The dispersion shown in Fig.~\ref{fig:outofplane} is obtained from the exact expression of the polarization operator. The small asymmetry, clearly visible in this numerical calculation, especially in the gapped branch, gives an additional broadening after averaging over domains of different chirality. Our results are valid for momenta $q=|{\bf k}\pm{\bf Q}| \lesssim \sqrt{x}$ and in the absence of disorder. In this range of momenta, disorder is expected to lead to some broadening of the neutron scattering peaks, as observed in elastic scattering, see Fig.~\ref{fig:structurefactor}. This broadening is getting stronger for momenta corresponding to the inverse separation between impurities, where spin waves are scattered off random impurities. In this region, the peaks are very broad and thus difficult to detect. However, we expect a distinct signal to reappear at higher momenta, reflecting the usual spin-wave dispersion $\omega=ck$. This reappearance is due to the nearly perfect antiferromagnetic alignment of nearest neighbor spins. The neutron scattering probability defined by the interaction~(\ref{n}) reads
\begin{eqnarray}
\label{i2}
&&I_{out}(\omega,{\bf q})\propto
-\frac{1}{4\pi}\text{Im} \ G_{out}(\omega,{\bf q}) \\
&&\approx \frac{1}{8\chi_{\perp}\omega}\left[ Z_{\bf q}\delta(\omega-\Omega_{1,{\bf q}})
+(1-Z_{\bf q})\delta(\omega-\Omega_{2,{\bf q}})\right] \ ,\nonumber
\end{eqnarray}
with the quasi-particle residue
\begin{equation*}
Z_{\bf q}=\frac{1}{1+\frac{c^2}{8g^2}\frac{q_b^2}{Q^2}} \ .
\end{equation*}
Since the upper branch has a practically vanishing spectral weight, it is difficult to observe experimentally. Compared to the in-plane spin waves excited in neutron scattering, the out-of-plane spectrum is very similar, especially around momenta ${\bf q}\to{\bf Q}$ where the energies go to zero. The difference can be best seen around momenta $q \approx 0$. However, one should remember that there is considerable broadening of the peaks due to disorder in this region. Interestingly, the ratio between the predicted intensities for the low-frequency in- and out-of-plane spin waves is $I_{out} : I_{in} = 2$, see Eqs.~(\ref{i1}) and (\ref{i2}). Such a difference should be detectable in experiments.

\section{Conclusion\label{sec:conclusion}}
To conclude, we have derived the low-energy effective field theory of the $t$-$J$ model in the limit of  small doping. Based on this description, we have calculated the spin-wave excitations in the disordered spin spiral state of La$_{2-x}$Sr$_x$CuO$_4$ (the spin-glass phase), which has a coplanar spin structure. For the in-plane spectrum, we have found the usual linear spin-wave dispersion, while the out-of-plane modes have been shown to exhibit non-trivial doping dependent features.

\acknowledgments
We are grateful to W.~Metzner, G.~Khaliullin, B.~Keimer, V.~Hinkov, M.~Silva Neto, A.~Katanin,
J.~Sirker, A.~Muramatsu, and I.~Affleck for valuable discussions. O.~P.~S. gratefully acknowledges  the support from the Alexander von Humboldt Foundation, and the hospitality of the Max-Planck-Institute for Solid State Research Stuttgart and Leipzig University.

\end{document}